\documentclass[%
reprint,
superscriptaddress,
showpacs,
amsmath,amssymb,
aps,
prx,
longbibliography,
]{revtex4-2}

\usepackage{psfrag,graphicx,epsfig,color}
\usepackage[usenames,dvipsnames,svgnames,table]{xcolor}
\usepackage{dcolumn}
\usepackage{bm}
\usepackage{rotating}
\usepackage{float}
\usepackage{lipsum}

\usepackage{natbib}

\def\re    {{R_\lambda}}
\def\uu {{\mathbf{u}}}
\def\xx {{\mathbf{x}}}
\def\rr {{\mathbf{r}}}
\def\kk {{\mathbf{k}}}
\def\ee {{\mathbf{e}}}
\def\ff {{\mathbf{f}}}

\def\ww {{\boldsymbol{\omega}}}
\def\zz {{\boldsymbol{\zeta}}}

\begin{document}


\title{
Twisting vortex lines regularize Navier-Stokes turbulence
}

\author{Dhawal Buaria}
\email[]{dhawal.buaria@ttu.edu}
\affiliation{Department of Mechanical Engineering,
Texas Tech University, Lubbock, TX 79409}
\affiliation{Max Planck Institute for Dynamics and Self-Organization, 37077 G\"ottingen, Germany}
\author{John M. Lawson}
\affiliation{Department of Aeronautics and Astronautics, University of Southampton, 
Southampton SO17 1BJ, UK}
\author{Michael Wilczek}
\affiliation{Theoretical Physics I, University of Bayreuth, 95440 Bayreuth, Germany}
\affiliation{Max Planck Institute for Dynamics and Self-Organization, 37077 G\"ottingen, Germany}

\date{\today}


\begin{abstract}


Fluid flows are
intrinsically characterized via the 
topology and dynamics of underlying vortex lines.
Turbulence in common fluids like water and air, 
mathematically described by the 
incompressible Navier-Stokes equations (INSE), engenders spontaneous 
self-stretching and twisting of vortex lines, 
generating a complex hierarchy of structures. 
While the INSE are routinely utilized to describe turbulence,
their regularity remains unproven;
the implicit assumption being that the self-stretching 
is ultimately regularized by viscosity,
preventing any singularities.
Here, we uncover an inviscid regularizing mechanism stemming from 
self-stretching itself, by analyzing the 
flow topology as perceived by an observer aligned with the vorticity 
vector undergoing amplification. 
While initially vorticity amplification occurs via increasing
twisting of vortex lines, a regularizing anti-twist
spontaneously emerges to prevent unbounded growth. 
By isolating a vortex we additionally demonstrate the 
genericity of this 
this self-regularizing anti-twist. 
Our work, directly linking dynamics of vortices 
to turbulence statistics, reveals how the Navier-Stokes 
dynamics avoids the development of singularities even without
the aid of viscosity. 


\end{abstract}

\maketitle

\section*{Introduction}

Vortices, constituting regions
of pronounced rotational coherence, are 
intrinsic to all fluid flows, both classical
and quantum, and various other physical
phenomena involving plasmas and electromagnetism
\cite{she_90, polanco21, kleckner2013, zwierlein, freilich2010, aharon}. 
From collisions of microscopic 
water droplets in clouds \cite{falkovich02} to global
circulation patterns on planets and stars \cite{dritschel},
and from nanodevices \cite{ahmed2012} 
to aerospace vehicles \cite{spalart1998},
they play a critical role in all natural and man-made
hydrodynamic phenomena.
Mathematically, vortices are characterized by
the vorticity vector $\ww = \nabla \times \uu$, where 
$\uu$ is the velocity field; thus, fluid flows can 
be inherently described as an ensemble of vortex lines,
akin to streamlines depicting the velocity field.
Their evolution is  
routinely described by the 
incompressible Navier-Stokes equations (INSE),
written in the vorticity form:
\begin{align}
\frac{D \ww}{D t} = \ww \cdot \nabla \uu + \nu \nabla^2 \ww  \ , 
\label{eq:vort}
\end{align}
where $D/Dt =\partial_t +  \uu \cdot \nabla $ is 
the material derivative, 
$\nu$ is the kinematic viscosity
and incompressibility implies 
$\nabla \cdot \uu = 0$, i.e.,
velocity is solenoidal.

\begin{figure*}
\begin{center}
\includegraphics[width=0.92\linewidth]{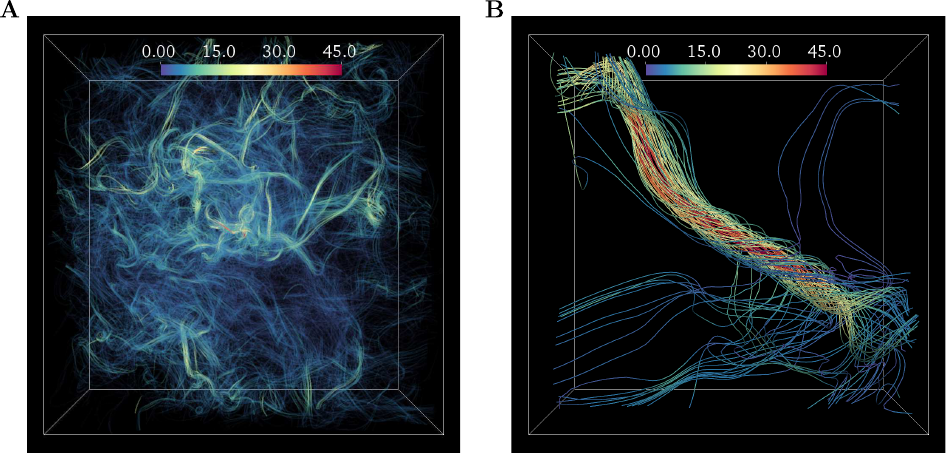} 
\end{center}
\caption{
{\bf Vortex lines in turbulence.}
a) The structure of vortex lines in instantaneous vorticity 
field from direct numerical simulation of turbulence at $\re=1300$.
The size of the domain shown is $(256 \eta)^3$, where
$\eta$ is the Kolmogorov length scale. The vortex lines
are color coded with vorticity magnitude
(as normalized by the rms). The visualization reveals bundling
of vortex lines around intense events, into well known
tube-like structures. 
b) A zoomed in view 
around the center of domain in panel a,
showing an individual vortex bundle around the 
intense event,
demonstrating conspicuous twisting
of vortex lines.
The size of the sub-domain in panel b is $(50 \eta)^3$.
}
\label{fig:1}
\end{figure*}

Describing a wide range of fluid dynamical phenomena,
the INSE are of central importance in 
science and engineering,
and are routinely utilized in 
a wide array of numerical
simulations \cite{pletcher}. 
However, from a 
mathematical standpoint, it is still unknown if they 
are well posed, i.e, whether solutions to INSE
always remain smooth or develop singularities
in finite time, which would preclude their widespread usability.
Consequently, their regularity problem has been recognized 
by the Clay Mathematics Institute as one of the 
Millennium Prize problems
\cite{Fefferman,doering2009}. 
The possibility of a finite-time singularity
arises from the nonlinear term $\ww\cdot\nabla \uu$  
in Eq.~\eqref{eq:vort},
which prescribes amplification
of vorticity via self-stretching of vortex lines 
(or vortex stretching), whereas the viscous
term acts to oppose this amplification 
\cite{tl72,Tsi2009,BBP2020}.
It follows that for a singularity
the nonlinear term must grow unbounded \cite{beale84}
which, if possible, can only happen
when the viscosity is sufficiently small \cite{doering2009}.
This corresponds to the turbulent regime,
a far-from-equilibrium state characterized by chaotic
multiscale fluctuations, leading
to intermittent generation of extreme vorticity
events and a highly complex spatiotemporal structure of
underlying vortex lines 
\cite{she_90,Jimenez93,Ishihara09,BPBY2019}.
Figure~\ref{fig:1} illustrates the structure 
of vortex lines in turbulence,
revealing prevalence of a hierarchy of coherent
vortex structures. 
In regions of intense vorticity, these structures
consist of bundles of high-amplitude vortex lines,
which display an inner twist, among other
topological properties.

The regularity problem for the INSE 
and intermittency in turbulent flows
are intimately connected
\citep{doering2009,Saw:2016,tao2019,BPB2020}. 
However, both problems
remain inherently challenging due to underlying mathematical
difficulties in deciphering the nonlinear
amplification term in Eq.~\eqref{eq:vort}.
It is worth noting that this term is 
closed in vorticity, since velocity can
be obtained from vorticity via the 
Biot-Savart integral over the flow domain:
\begin{align}
\uu(\xx,t) = \frac{1}{4\pi} 
\int_{\rr}
\frac{\rr}{|\rr|^3} 
\times \ww (\xx + \rr, t)   \ d\rr \ . 
\label{eq:biot}
\end{align}
This nonlocal integral thus couples all the scales in the flow,
but also renders Eq.~\eqref{eq:vort}  analytically
intractable, posing a serious challenge in
addressing the nonlinearity \cite{doering2009}.
Some understanding of this nonlocality can be 
obtained by evaluating the Biot-Savart integral
from numerical simulations of INSE, and analyzing
its behavior in neighborhood of intense vorticity, 
see e.g. \cite{ham_pof08,BBP2020,BP2021}.
From a geometrical standpoint, 
a singularity is engendered by
untamed stretching and twisting of vortex lines,
ultimately rendering them non-smooth 
\cite{constantin1993,constantin1996}
and simultaneously leading to 
unbounded growth of the nonlinear term.
While recent works \cite{BPB2020,BP2021}
have argued against unbounded growth,
they do not provide
the necessary geometrical and structural 
understanding of the flow 
around intense vorticity or any dynamical
information on evolution of vortex lines as a whole --
which is essential to understand the behavior of the nonlinearity
in the INSE.
Indeed, such a task is inherently challenging
due to the complex hierarchy of vortex structures 
prevalent in the turbulent regime,
and often  instead restricted to simplified
vortex flows with specific initial conditions
\cite{kerr:1993,luo14}.

In this work, 
by taking the perspective of an observer
aligned with vorticity undergoing amplification,
we isolate
the topology and dynamics of vortex lines
around it to highlight the physical mechanism
underpinning the regularity of Navier-Stokes turbulence.
In such a conditional reference frame, 
vorticity amplification necessitates that 
vortex lines possess a degree of twist around 
the observer \cite{novikov93jfr,mui1996}.
However, we demonstrate that as vorticity is increasingly amplified, 
an anti-twist is spontaneously generated within
the vortex core which attenuates further amplification.
The robustness of this observation is further
verified by isolating an individual vortex, which 
is then evolved using the INSE, once for finite
viscosity and once with viscosity explicitly set to zero
(effectively simulating Euler equations).
In both cases, the same qualitative behavior is
obtained, with the temporal dynamics
demonstrating the emergence of the self-regularizing
anti-twist as vorticity is amplified.
This mechanism
provides a natural explanation for the recently identified
self-attenuation of extreme vorticity events in turbulence
\cite{BPB2020}, and 
elucidates how the nonlinear dynamics of Navier-Stokes equations
preclude formation of any singularities.

\section*{Results}

\paragraph*{\bf CAV framework:}
The change in reference frame is accomplished by
utilizing the so-called conditionally averaged vorticity 
(CAV) field
\begin{align}
\overline{\ww}(\rr,\zz) = \langle \ww (\xx + \rr, t) \big | \zz = \ww (\xx,t )  \rangle \ , 
\label{eq:cav0}
\end{align}
which essentially captures how an observer
aligned with vorticity in a particular state $\zz = \ww(\xx, t)$ at 
any location $\xx$ (and time $t$),
perceives the  neighboring vorticity field 
$\ww(\xx+\rr, t)$ at any distance $\rr$ from $\xx$. 
The average $\langle \cdot \rangle$ is taken over
multiple independent realizations (we elaborate further on this later). 
The CAV field was originally proposed by Novikov \cite{novikov93jfr}
and has been utilized in the context of 
turbulence theory and closure modeling 
\cite{novikov93jfr,mui1996,wilczek_thesis,Friedrich:2012}. 
However, by aligning with the 
vorticity vector $\ww(\xx,t)$ undergoing amplification,
the CAV field also allows us to analyze the
flow topology in its neighborhood. 
To this end, we consider
the CAV field in cylindrical polar coordinates
$(\rho, \theta, z)$:
\begin{align}
\overline{\ww}(\rr,\zz) =
\overline{\omega}_z \ee_z + \overline{\omega}_\rho \ee_\rho 
+ \overline{\omega}_\theta \ee_\theta \ , 
\label{eq:cav}
\end{align}
where the 
orthogonal unit vectors 
are given as:
$\ee_z = \hat{\ww} = \zz/|\zz|$, 
$\ee_\rho = (\hat{\rr} - \gamma \ee_z)/(1-\gamma^2)^{1/2}$,
$\ee_\theta = \ee_z \times \ee_\rho$, with
$\hat{\rr} = \rr/|\rr|$ and $\gamma=\hat{\rr} \cdot \ee_z$.
The vorticity condition $\zz = \ww(\xx, t)$
is always centered at the origin $\rr=\mathbf{0}$ in this frame and
points along $\ee_z$. 
It simply follows that the coordinates
$\rho = \rr \cdot \ee_\rho = r (1-\gamma^2)^{1/2}$, and 
$z= \rr \cdot \ee_z = r\gamma$,
with  $r = |\rr| = (\rho^2 + z^2)^{1/2}$.
For convenience, we also define
the enstrophy $\Omega = |\ww(\xx)|^2$,
which simply quantifies the magnitude
of the vorticity undergoing amplification.

In general, the CAV field is a function of vectors
$\rr$ and $\zz$, i.e., six variables; 
however, using statistical isotropy,
the dependence is reduced to three scalar variables $\Omega$, $r$ and $\gamma$
\cite{mui1996}, 
i.e., the magnitudes of $\rr$ and $\zz$ and the relative
alignment between them. 
Thus,  $\overline{\omega}_{z, \rho, \theta} = 
\overline{\omega}_{z, \rho, \theta} (\Omega, r, \gamma)$
or alternatively, in the cylindrical polar coordinates:
 $\overline{\omega}_{z, \rho, \theta} = 
\overline{\omega}_{z, \rho, \theta} (\Omega, \rho, z)$,
which implies that the CAV field is axisymmetric,
naturally justifying the use of these coordinates.

Similar to the CAV field, one can also extract the conditional
velocity field around $\xx$ using
the same procedure, i.e., $\langle \uu(\xx+\rr, t) | \zz \rangle$. 
However, the velocity field can also be obtained
from the CAV field using the Biot-Savart integral in Eq.~\eqref{eq:biot}. 
The nonlinear 
amplification term in Eq.~\eqref{eq:vort}  
can then be obtained
in terms of the CAV field \cite{novikov93jfr,mui1996}.
Since we are predominantly concerned with 
vorticity magnitude for a potential singularity, 
we first take the dot product
of Eq.~\eqref{eq:vort} with $\ww$, which
gives an equation for $\Omega$.
The nonlinear amplification term is then given as
(see Methods for derivation):
\begin{align}
\langle (\hat{\ww} \cdot \nabla \uu) \cdot \hat{\ww} | \Omega \rangle =
 \int_0^\infty \int_0^\infty 
\frac{3 \rho^2 z }{r^5} \ \overline{\omega}_\theta(\Omega, \rho, z) \ d\rho \ dz \ .
\label{eq:wsw}
\end{align}
Remarkably, the nonlinear stretching
of vorticity can be solely written in terms
of $\overline{\omega}_\theta$, which represents the twist
associated with the CAV field. 
Note that this connection was already realized in earlier works
\cite{novikov93jfr,mui1996}, albeit formulated
in spectral-space;  whereas here we 
have reformulated it in physical space.
which is more intuitive and useful for the subsequent discussion.
This expression is also linear
in $\overline{\omega}_\theta$, which provides remarkable simplification
in analyzing the nonlinearity when using the CAV framework. 
The full nonlinear
term can be obtained as
$\langle (\ww \cdot \nabla \uu) \cdot \ww |\Omega \rangle 
= \Omega  \langle (\hat{\ww} \cdot \nabla \uu) \cdot \hat{\ww} |\Omega \rangle $.
We have also  utilized the fact that 
$\overline{\omega}_\theta(z) = -\overline{\omega}_\theta(-z)$,
i.e., $\overline{\omega}_\theta$ is an odd-function of $z$, which simply follows
from rotational symmetry of the CAV field.



From Eq.~\eqref{eq:wsw}, it is evident that
for the nonlinear stretching to be positive
the integral on the right-hand side must be positive,
leading to the expectation that 
$\overline{\omega}_\theta(z) > 0$ for $z>0$
(and $\overline{\omega}_\theta(z) <0$ for $z<0$ from 
symmetry). Thus, we anticipate a positive
twist of the CAV, which increases in strength
as the nonlinear amplification increases 
to enable vorticity amplification. 
We will indeed demonstrate this result in the next
section. But more remarkably, we will demonstrate
that when vorticity magnitude is sufficiently
large, a negative anti-twist emerges, i.e.,
$\overline{\omega}_\theta (z)<0$ for some $z>0$,
in conjunction with
positive twist in the background,
which then attenuates further vorticity amplification.
It is worth emphasizing that the twist component 
$\overline{\omega}_\theta$
explicitly relates to the nonlinear term only, i.e.,
it does not contribute to the viscous
term \cite{mui1996,wilczek_thesis}. Thus, the emergence of anti-twist
explicitly amounts to an inviscid regularizing mechanism
originating from the nonlinearity itself.

\begin{figure*}
\begin{center}
\includegraphics[width=0.92\linewidth]{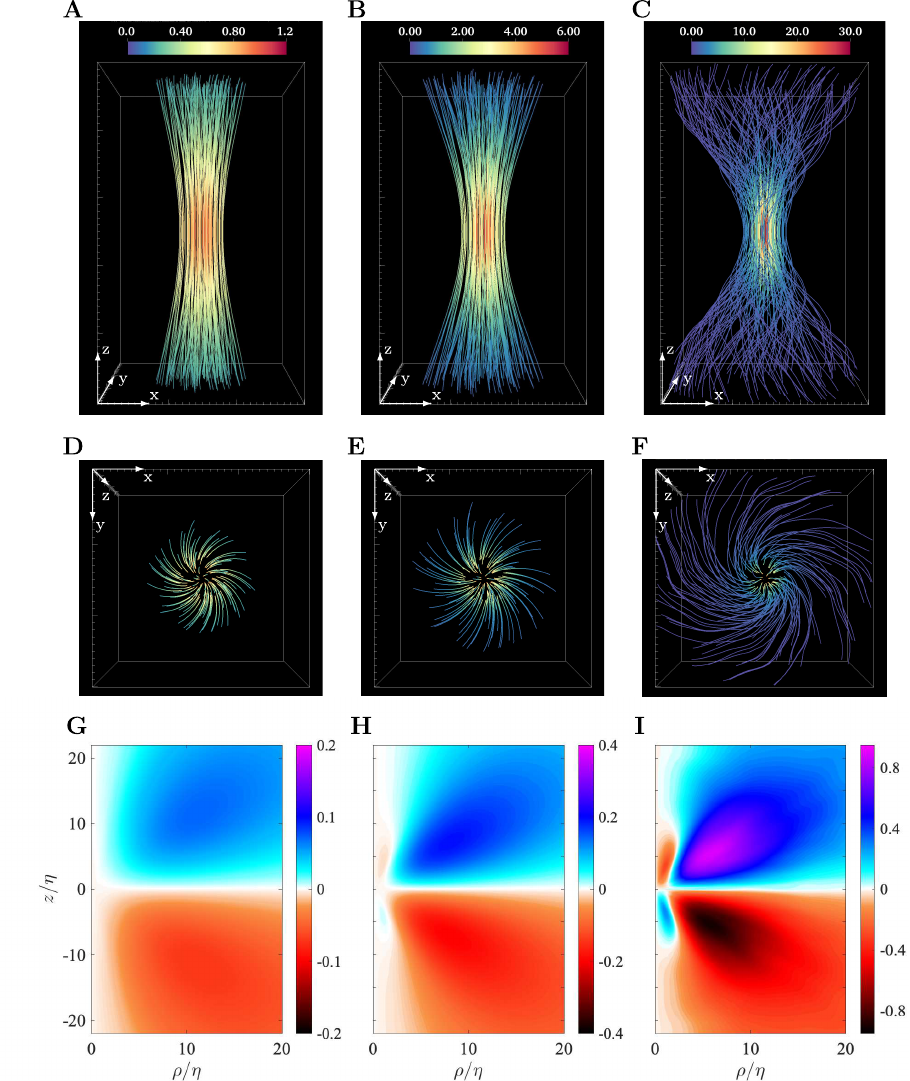}
\end{center}
\caption{
{\bf Conditionally averaged vorticity (CAV) field 
obtained from DNS}.
a-c) Vortex lines showing the structure
of the CAV field for increasing
conditioning magnitudes of 
$\Omega/\langle \Omega \rangle = 1, 30, 1000$, respectively. 
The domain size in each case is $30 \eta \times 30 \eta \times 50 \eta$. 
Note, the condition vorticity $\zz$ is at the center 
of the domain (corresponding to $\rr=\mathbf{0}$). 
d-f) top view of the domain in a-c highlighting the increasing 
twisting of vortex lines with the conditioning magnitude,
and  g-i)  The twist component 
$\overline{\omega}_\theta$ 
of the CAV field, respectively from 
axisymmetric planar cuts in a-c. As the conditioning value increases, 
a clear anti-twist emerges in the center of the CAV field.
(The components 
$\overline{\omega}_z$ and $\overline{\omega}_\rho$ 
are shown in the supplementary Fig.~1 for completeness).
}
\label{fig:cav1}
\end{figure*}

\paragraph*{\bf Data:} 
To obtain the CAV field and associated
quantities, we utilize data from both 
laboratory experiments and a large database 
generated via direct numerical simulations (DNS) of the INSE. 
The laboratory experiments correspond
to turbulence measurements in a von-K{\'a}rm{\'a}n swirling 
water flow \cite{Lawson2019}. The measurements
are made in a small $1 \ \textrm{cm}^3$ volume near the mean-field stagnation point
using scanning particle image velocimetry 
(PIV), allowing us to obtain the full 3D velocity
(and vorticity) field \cite{Lawson2014}.
The simulations correspond to the canonical setup
of forced isotropic turbulence in a periodic domain
and are performed using the well known Fourier pseudo-spectral
methods, allowing us to obtain any quantity of interest with
highest accuracy practicable \cite{mm98, Ishihara09}. 
The intensity of turbulence in both experiments and DNS
is measured using the Taylor-scale Reynolds number $\re \equiv U \lambda/\nu$,
where $U$ is the {\em rms} of velocity and $\lambda$ is the Taylor length 
scale. Note that  $\re \sim Re^{1/2}$ \cite{Frisch95}, 
where $Re \equiv UL/\nu$ is the 
large scale Reynolds number, with $L$ being the characteristic large scale. 
In experiments, $\re\approx200$, whereas in DNS 
$\re$ is varied from $140$ to $1300$, on some of the 
largest grid sizes currently
feasible in turbulence simulations \cite{BPB2020, BP2022, BS_PRL_2022, BS_PRL_2023}.
In both cases, special attention is given to resolve the 
small scales and hence the intense vorticity events 
accurately \cite{BPBY2019, BP2022},
keeping the spatial resolution smaller than or equal to
the Kolmogorov length scale $\eta$, which characterizes the 
cut-off scale at which viscosity regularizes the flow \cite{Frisch95}.
Full details about both experiments and DNS are provided
in the Methods section. The precise details about
the extraction of the CAV field from the data are also provided there.

\paragraph*{\bf Structure of CAV field:}
Our first key result is shown in 
Fig.~\ref{fig:cav1}, which illustrates the 
structure of the CAV field with increasing
magnitude of enstrophy. The panels a-c
show the visualization of the vortex lines in the neighborhood
of the conditioned vorticity, which is always
at the center of the domain and pointing upwards
(along the $z$-axis).
Each visualization reveals an
elongated axisymmetric vortex with 
conspicuous twisting of vortex lines, which is 
concentrated near the center 
of the vortex and fans out along the elongated direction.
The twisting also gets stronger and more prominent
as the magnitude of the conditioned vorticity
increases, which essentially captures
the increased stretching required to generate
more intense vorticity \cite{BBP2020} -- as also dictated by Eq.~\eqref{eq:wsw}.
To further highlight this aspect,
panels d-f show the vortex structures in panels a-c,
as viewed bottom to top (along the $z$-axis). 
The increasing span of the vortex lines
reiterates the increase in twist
with the conditioning magnitude.

\begin{figure}
\begin{center}
\includegraphics[width=0.98\linewidth]{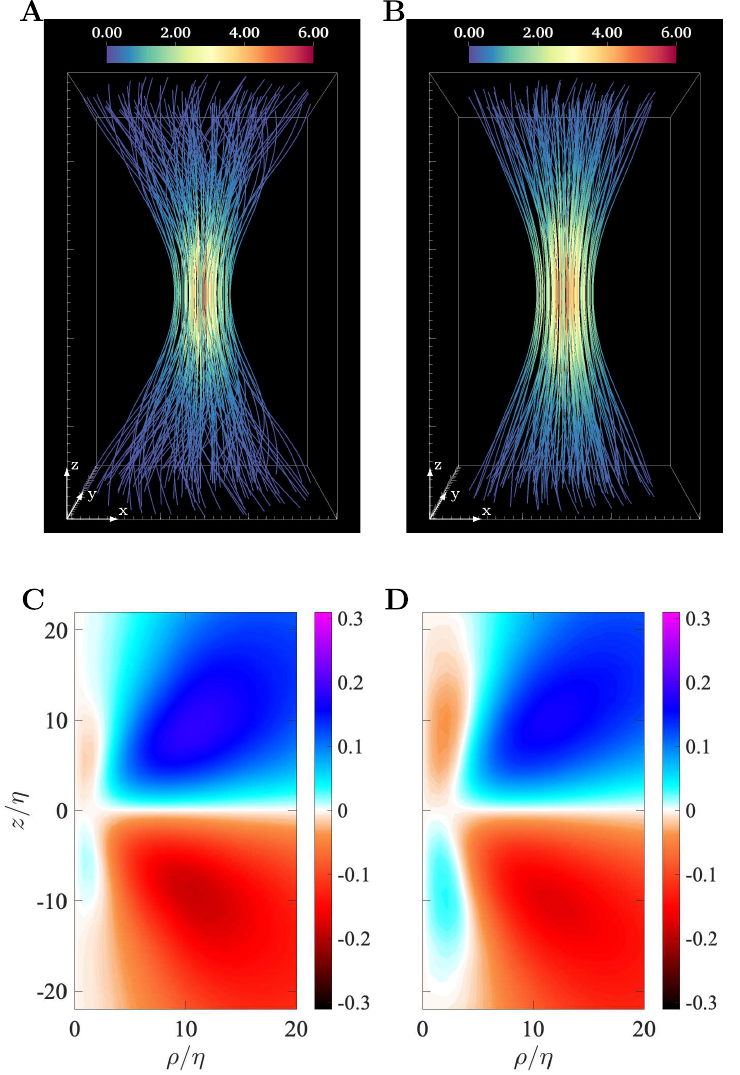} 
\end{center}
\caption{
{\bf Comparing CAV fields from DNS
and experiments} 
a-b) Vortex lines showing the structure of the CAV field
in DNS (panel a) and experiments (panel b),
c-d) The twist component $\overline{\omega}_\theta$ 
of the CAV field, respectively from 
axisymmetric planar cuts in a-b. 
The data in both cases corresponds to $\re = 200$
and conditional  $\Omega/\langle\Omega\rangle = 30$.
(The components $\overline{\omega}_z$ and 
$\overline{\omega}_\rho$ 
are shown in the supplementary Fig.~2 for completeness;
whereas the supplementary Fig.~3 presents a comparison
at $\Omega/\langle\Omega\rangle = 1$, which does not have the anti-twist).
}
\label{fig:dns_exp}
\end{figure}

\begin{figure*}
\begin{center}
\includegraphics[width=0.99\linewidth]{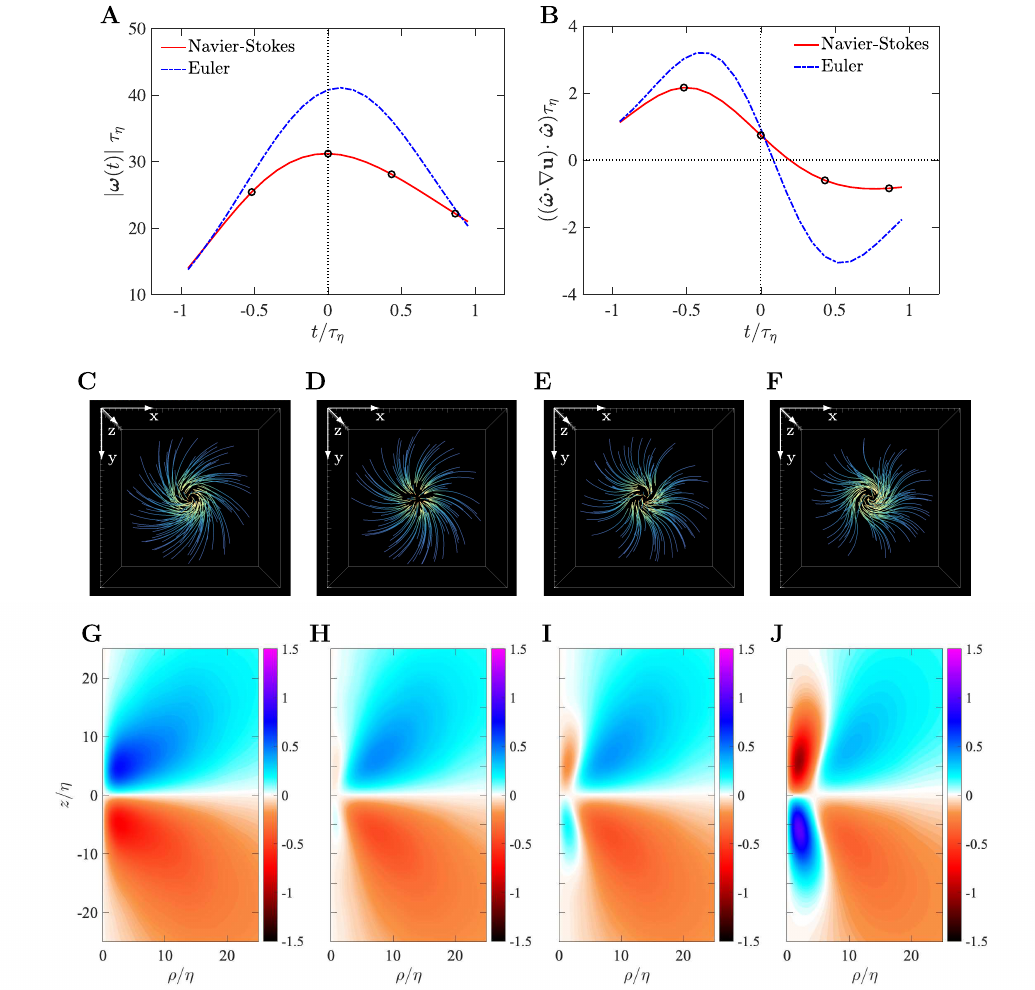} 
\end{center}
\caption{
{\bf Temporal dynamics of an isolated vortex structure}.
a) Evolution of vorticity magnitude
at $\rr = 0$, as a function
of non-dimensional time $t/\tau_\eta$, 
where $\tau_\eta = \langle \Omega \rangle ^{-1/2}$ 
is the Kolmogorov time scale. 
The Euler case corresponds to the result 
when viscosity is set to zero.
b) Evolution of total stretching, also at $\rr = 0$. 
c-f) the structure of vortex lines (top view)
corresponding to the four time instants marked in 
panels a and b. Initially the vortex lines only have a positive
twist, but later an anti-twist develops in the center.
g-j) The twist component $\overline{\omega}_\theta$
also at the same four time instants, again showing
the emergence of anti-twist. 
}
\label{fig:cav_lag}
\end{figure*}

As noted earlier, the observation of a (positive) twist is 
expected on the grounds that vortex stretching is positive,
and the observed structure of the CAV field is consistent with 
earlier results of \cite{mui1996, wilczek_thesis} in that regard.
However, a striking previously overlooked feature is the anti-twist 
region that emerges for intermediate enstrophy condition,
and grows larger as enstrophy condition intensifies. 
While this aspect it difficult to observe from vortex lines,
it is readily evident from 
the azimuthal component $\overline{\omega}_\theta$ of the CAV
shown in Fig.~\ref{fig:cav1}g-i,
corresponding to planar cuts through the mid plane 
in both radial and axial directions from Fig.~\ref{fig:cav1}a-f. 
Evidently, the magnitude of $\overline{\omega}_\theta$  is overall
stronger when the CAV field for stronger enstrophy is 
considered.
Since net positive stretching is
required to generate intense vorticity,
it also follows from Eq.~\eqref{eq:wsw} that
$\overline{\omega}_\theta$ is mostly positive for $z>0$ (and 
also negative for  $z<0$ due to the odd symmetry 
$\overline{\omega}_\theta (\rho, z) = - \overline{\omega}_\theta(\rho, -z)$).
The negative anti-twist close to the CAV center opposes the 
positive background twist responsible for amplification of vorticity
(and generation of enstrophy). 

It is worth noting that the anti-twist observed in the CAV field
also provides an explanation for the self-attenuation mechanism 
recently identified in \cite{BPB2020}. 
In this work, it was observed that in regions of very intense vorticity,
the locally induced stretching is in fact negative and counteracts
the net positive stretching. 
Our results here identify the anti-twist in the CAV as the key feature in 
the vorticity field that implies negative local stretching, essentially
explaining the mechanism in \cite{BPB2020}.

Related to that, the twisted vortex structures identified in Fig.~\ref{fig:cav1}a-c  
bear noticeable similarity
to vortex tubes directly extracted from the instantaneous flow in Fig.~\ref{fig:1}. 
This is somewhat expected, since the CAV field essentially represents the typical 
local structures of the flow. In this context it is worth mentioning that the vortex 
tubes in turbulence are often represented by axisymmetric Burgers vortices \cite{Burgers48}. 
However, Burgers vortices do not have any twisting and are incapable of 
self-amplification; instead, their enstrophy generation is facilitated 
via a constant background strain field.
A similar observation was also made in \cite{BPB2020}, which identified
local preferential alignment between velocity and vorticity, i.e., 
net positive helicity, in regions of intense vorticity. 
Once again, the twisting of vortex lines in the CAV field provides a natural
explanation for this.

To establish that this anti-twist is a generic feature of
turbulence, we present a comparison
of the DNS results to those obtained from 
a turbulent flow in laboratory experiments under very different 
conditions (see Methods for details).  
Figure~\ref{fig:dns_exp} shows the 
vortex lines and the $\overline{\omega}_\theta$ corresponding
to most intense vorticity available from
DNS and experiments. The comparison is done at $\re=200$,
which is smaller than utilized in
Fig.~\ref{fig:cav1}
(since higher $\re$ are not available in experiments). 
Nevertheless, as evident from Fig.~\ref{fig:dns_exp}, 
the development of the anti-twist at
large conditional vorticity is an inherent property
of turbulence. 
Is is worth noting that 
the emergence of the anti-twist occurs
at a lower enstrophy condition at lower Reynolds number. 
This is essentially because of intermittency, which dictates
that the extremeness of an extreme enstrophy event
(with respect to the mean value) increases
with Reynolds number \cite{BPBY2019, BP2022}.
A more detailed comparison between 
CAV fields
from experiments and DNS is shown in the 
Supplementary Material (in Figs.~2-3), comparing
all the components of CAV, for
two different vorticity conditions.

\paragraph*{\bf Dynamics of an isolated vortex:}
The CAV results shown in Figs.~\ref{fig:cav1}-\ref{fig:dns_exp}
demonstrate the existence of an anti-twist region
in the vortex core when vorticity is amplified to large
magnitudes. However, they do not illustrate
the dynamical origin of such an anti-twist. 
Since self-amplification of vorticity
is an inherent characteristic of turbulence
\cite{Betchov56, Tsi2009, BBP2020},
it follows from Eq.~\eqref{eq:wsw} 
that the vortex
lines develop a twist to enable
this amplification. In the following, we 
demonstrate
that the nonlinearity also 
inherently encodes the development of the
countering anti-twist as a regularizing mechanism.

To this end, we perform a simple numerical experiment: 
We consider an isolated
vortex structure extracted from the
CAV field which, as an essential feature, only exhibits 
the regular twisting
of vortex lines is present (see Methods). 
This initial condition corresponds to 
a vortex structure with positive twisting everywhere,
enabling self-amplification of vorticity. 
Essentially, we can write 
$\overline{\omega}_\theta > 0$ for all $z>0$ and
$\overline{\omega}_\theta < 0$ for all $z<0$. 
Due to the axisymmetry of 
the CAV field, the vorticity vector undergoing
maximum amplification is always at the center
of the vortex core ($\rr= {\bf 0}$) and pointed towards positive $z$-axis.
With this state as the initial condition,
we then evolve it using the 
Navier-Stokes equations and also
the inviscid Euler equations \cite{gibbon08} (which  
corresponds to setting the viscosity 
to zero in the numerical simulation). 
In both cases, we demonstrate
that an anti-twist naturally develops
in the vortex core as vorticity is sufficiently amplified. 
It is worth noting that this approach is similar
in essence to considering simplified
vortex flows with arbitrary initial conditions, 
see e.g. \cite{kerr:1993, luo14, elgindi2021}, which are
often utilized to search for singularities.
The key innovation of our approach is that the initial 
condition is directly obtained from the CAV field
and is therefore representative of turbulence.

The temporal evolution of the isolated vortex structure,
is shown in Fig.~\ref{fig:cav_lag}. 
Fig.~\ref{fig:cav_lag}a shows the temporal evolution
of the vorticity magnitude at
the center of the core.
The initial state is the same for both Navier-Stokes
and Euler simulations and the initial time is taken
to be negative for
convenience, with $t=0$ corresponding
to a later time when maximum amplitude is reached 
in the Navier-Stokes simulation.
We first focus on the Navier-stokes simulation and will come
back to the Euler case.
For $t>0$, the vorticity magnitude starts decreasing. 
The evolution of the flow, via vortex lines, is shown in
Fig.~\ref{fig:cav_lag}c-f, corresponding
to the times marked in Fig.~\ref{fig:cav_lag}a. 
For convenience, the top view 
is shown (similar to Fig.~\ref{fig:cav1}d-f). 
It can be seen that initially
field (in panel c), the vortex lines all twist clockwise,
when going outwards from origin. 
However, as time progresses, 
an anti-twist emanates
from the vortex core (with lines going counterclockwise
and essentially
acting to oppose the regular clockwise twist in the background).
To better quantify this, Fig.~\ref{fig:cav_lag}g-j
shows the corresponding contour fields of the twisting component
$\overline{\omega}_\theta$ for the same times. 
Once again, we notice 
(in panel g) that the twisting 
is positive everywhere at the initial time.
However, in panel h, we can clearly see that
as vorticity is self-amplifying, an anti-twist 
is also simultaneously produced.
In panels i-j, this counter twist region
grows bigger, corresponding
to further attenuation of vorticity. 


In Fig.~\ref{fig:cav_lag}b, the corresponding
temporal evolution of the total stretching,
acting on the vorticity at $\rr = \bf 0$ is shown
(as also defined in Eq.~\eqref{eq:wsw}). 
We notice that initially the stretching increases
with time,  enabling vorticity amplification
(as shown in Fig.~\ref{fig:cav_lag}a).
However, even before maximum of vorticity magnitude
is reached (at $t=0$), the stretching starts to decrease,
corresponding to the emergence of the anti-twist. 
As the anti-twist region grows, the net stretching
eventually becomes negative.

As mentioned earlier, we also consider
the Euler simulation, which corresponds
to the Navier-Stokes case with viscosity 
set to zero. 
The behavior of the Euler simulation
is shown in both Fig.~\ref{fig:cav_lag}a and b. 
Essentially the same qualitative behavior is obtained
as the Navier-stokes case, with some minor quantitative
differences.
For instance, the maximum amplitude of vorticity is
slightly higher for the Euler run and
reached at $t>0$, essentially
because viscosity still plays a noticeable role
in attenuating vorticity (which is expected to get smaller
at very high Reynolds numbers, which are currently inaccessible
in DNS). 
Likewise, although not shown, the same qualitative behavior
is obtained for the Euler case, as the Navier-Stokes
case shown in Fig.~\ref{fig:cav_lag}c-j.

Based on the results shown in Fig.~\ref{fig:cav_lag},
the following picture emerges. 
The initial condition sets up 
for self-amplification of the isolated vortex
under its own twist. However, as 
it undergoes amplification, the anti-twist emerges from the core,
leading to attenuation
of the vortex. 
Thus, it follows that
this self-attenuation mechanism
stems from 
nonlinear dynamics of the equations,
implying that they 
already encode a nonlinear
mechanism to regularize extreme events.
While this conclusion can be drawn from the Navier-Stokes
run itself, the Euler simulation further reinforces it. 
The presence of such a mechanism was  
recently hinted in \cite{BPB2020}. 
Our results here establish the anti-twist in the CAV as 
the key statistical feature in the vorticity field that 
enables this self-regularization.

\section*{Discussion}

In this work, we have utilized both well resolved numerical simulations and 
laboratory experiments to analyze the behavior of vortex lines in 
the neighborhood of intense vorticity regions, which are 
signatures of potential singularities of Navier-Stokes turbulence. 
We employ the unique CAV framework \cite{novikov93jfr,mui1996}, which 
allows us to change the reference frame and align the observer with vorticity 
undergoing amplification. In this frame, the 
nonlinear self-stretching of vortex lines, the only possible source
of singularity, can be solely expressed in
terms of the twist component of the CAV field. 
Since vorticity amplification is an intrinsic property of turbulence,
a positive twist in the CAV field is expected and observed,
also in accordance with previous works \cite{mui1996}. 
However, we find that the CAV field remarkably generates a 
negative anti-twist which locally attenuates intense vorticity.
By isolating a vortex structure which initially possesses only 
positive twist, we demonstrate the spontaneous emergence of
the anti-twist when evolved under both Navier-Stokes and inviscid Euler dynamics. 
The development of the anti-twist, encoded in the nonlinearity,
therefore provides an inviscid means to prevent singular behavior
in Navier-Stokes equations \cite{Fefferman}. 

The results established within the CAV framework in the current work
relate to various recent observations in a similar context.
In ref.~\cite{BPB2020, BP2022}, a self-attenuation mechanism was identified
in the nonlinearity of Navier-Stokes, by analyzing the nonlocal coupling
between vorticity and strain fields. The emergence of the anti-twist
in vortex lines to attenuate amplification essentially
provides an explanation for the self-attenuation mechanism.
In this regard, it also would be interesting to understand the results
obtained here (using the CAV framework), in the more traditionally utilized
eigenframe of strain, particularly
by analyzing 
the dynamics of alignments between vorticity and eigenvectors
of strain, together with the sign of the intermediate eigenvalue 
\cite{Ashurst87, BBP2020}. 
Likewise, it could also be useful to relate the behavior of the CAV
field with that of the pressure field \cite{BPB2022, BP2023}.

It is also worth considering 
how the CAV framework is related to other statistical 
approaches to turbulence. The
CAV field also naturally relates to the 
instanton framework for the vorticity in three-dimensional turbulence,
which aims at capturing the statistics of extreme events \cite{grafke2015}.
Recent instanton calculations, analyzing extreme vorticity events
in turbulence, also remarkably showed signs of 
an emergent anti-twist \cite{grafke2015,schorlepp22prsa},
motivating further work on the implications of this feature 
for extreme-event statistics.

While the emergence of the self-regularizing anti-twist
ensures smoothness of vortex lines, an important future task
would be to leverage this physical mechanism 
to  establish rigorous bounds on
vorticity amplification \cite{constantin1996, doering2009}. 
In this regard, earlier works on core dynamics
of an isolated vortex column \cite{melander94} 
bear noticeable similarity to our own analysis presented here.
In \cite{melander94}, studying an isolated axisymmetric vortex
column---albeit in an idealized and simplified setting compared to the 
CAV field---the authors identified a differential rotation
mechanism arising due to variations in 
vortex core size, which naturally enables
twisting of vortex lines. Such variations in core size 
are also obtained in vortex structures identified using the CAV field,
and as discussed earlier, are notably absent in Burgers' vortices.
Taken together with the current results, this motivates further work towards 
a comprehensive understanding of vortex structures in turbulence, 
their potentially self-regularizing dynamics
(especially in connection to the Navier-Stokes regularity problem), 
and ultimately towards establishing a statistical description 
of turbulence.

Finally, it is worth emphasizing that although the current work
focuses on incompressible fluid turbulence, vortex lines are prevalent in 
various other phenomena, such as quantum turbulence,
plasmas and electromagnetism. In this context,
the unique CAV framework and insights derived from it 
offer a unique approach to analyzing these 
diverse and challenging problems. 



\section*{Materials and Methods}

\paragraph*{\bf Direct numerical simulations.}
The numerical data are obtained
through direct numerical simulations (DNS)
of the incompressible Navier-Stokes equations (INSE)
\begin{align}
\partial \uu/\partial t + \uu \cdot \nabla \uu = -\nabla P  + \nu \nabla^2 \uu + \ff \ ,
\end{align}
where $\uu$ is the divergence-free velocity field ($\nabla \cdot \uu = 0$), 
$P$ is the kinematic pressure, 
$\nu$ is the kinematic viscosity,
and $\ff$ corresponds to a large-scale forcing used to maintain a
statistically stationary state \cite{EP88}. 
The simulations correspond to the canonical setup
of homogeneous and isotropic turbulence,
with periodic boundary conditions on a cubic domain,
which is ideal for studying small scales and hence 
extreme events at highest Reynolds numbers 
possible \cite{Ishihara09}. The domain length in each direction
is $L_0 = 2\pi$, and discretized using $N$ points
with uniform grid spacing $\Delta x = L_0/N$. 
We solve the equations using a massively parallelized 
version of the well-known
Fourier pseudo-spectral algorithm of Rogallo (1981) \cite{Rogallo}
and the resulting aliasing errors are controlled by 
a combination of grid shifting and spherical truncation \cite{PattOrs71}.
Whereas for time integration, 
we use explicit second-order Runge-Kutta, with the time step $\Delta t$ 
subject to the Courant number ($C$)
constraint for numerical stability, i.e.,
$\Delta t = C \Delta x/ ||\uu||_\infty$ (where
$||\cdot||_\infty$ is the $L^\infty$ norm).

\begin{table}[h]
\centering
    \begin{tabular}{cccccc}
\hline
    $\re$   & $N^3$    & $k_{max}\eta$ & $T_E/\tau_K$ & $T_{sim}$ & $N_s$  \\
\hline
    140 & $1024^3$ & 5.82 & 16.0 & 6.5$T_E$ &  24 \\
    200 & $1024^3$ & 3.72 & 24.8 & 6.0$T_E$ &  24 \\
    240 & $2048^3$ & 5.70 & 30.3 & 6.0$T_E$ &  24 \\
    390 & $4096^3$ & 5.81 & 48.4 & 4.0$T_E$ &  28 \\
    650 & $8192^3$ & 5.65 & 74.4 & 2.0$T_E$ &  35 \\
   1300 & $12288^3$ & 2.95 & 147.4 & 20$\tau_K$ &  18 \\
\hline
    \end{tabular}
\caption{Simulation parameters for the DNS runs
used in the current work: 
the Taylor-scale Reynolds number ($\re$),
the number of grid points ($N^3$),
spatial resolution ($k_{max}\eta$), 
ratio of large-eddy turnover time ($T_E$)
to Kolmogorov time scale ($\tau_K$),
length of simulation ($T_{sim}$) in statistically stationary state
and the number of instantaneous snapshots ($N_s$) 
used for each run to obtain the statistics.
}
\label{tab:param}
\end{table}


The DNS database along with other
simulation parameters is shown in Table~\ref{tab:param}
and is the same as utilized in 
numerous recent works focused on small scale intermittency 
in turbulence,
see e.g. \cite{BBP2020, BS2022, BPB2020, BP2021, BPB2022, BS_PRL_2022, BS2023},
which establish the reliability and veracity of the data.
The Taylor-scale Reynolds number $\re$ is in the range $140-1300$.
To precisely compare with experiments (described next), 
we have additionally performed a new simulation at $\re=200$,
which was not reported before. 
The small-scale resolution in pseudo-spectral DNS 
is given by the parameter
$k_\text{max}\eta$, where $k_\text{max}=\sqrt{2}N/3$ is the maximum resolved
wavenumber on a $N^3$ grid and $\eta$ is the Kolmogorov length scale.
Equivalently, one can use the ratio $\Delta x/\eta$ ($\approx 2.96/k_\text{max}\eta$).
For all of our runs, we have very high spatial resolution,
going up to $k_\text{max}\eta\approx6$, 
to appropriately resolve the extreme events 
\cite{BPBY2019, BP2022}. 
This resolution can be compared
to the one used in comparable numerical investigations of turbulence
at high Reynolds numbers, which are mostly in the range
$1 \le k_\text{max} \eta \le 1.5$ \cite{Ishihara09,BSY.2015} -- which do not
resolve the extreme events adequately \cite{BPBY2019}.


\paragraph*{\bf Experiments.}
The experimental data constitute $2 \times 10^5$ statistically independent snapshots of 
the turbulent velocity field in a $1~\textrm{cm}^3$ measurement volume near the 
mean-field stagnation point of a von-K{\'a}rm{\'a}n swirling 
water flow, at $\re \approx 200$  \citep{Lawson2019}.
The flow facility consists of a $48~\textrm{cm}$ inner diameter, $58~\textrm{cm}$ 
tall stainless steel cylinder with internal baffles and filled with de-ionized water 
maintained at $21.2\pm 0.5^\circ\textrm{C}$ and agitated by two $25~\textrm{cm}$ 
diameter counter-rotating impellers rotating at $0.2~\textrm{Hz}$ \citep{Xu2007}.
Near the geometric centre of this vessel, homogeneous and 
axisymmetric turbulennce 
is generated with large-eddy length scale $L \equiv u'^3 / \epsilon \approx 77~\textrm{mm}$ 
and Kolmogorov length scale $\eta \approx 210~\mu\textrm{m}$.
The 3D velocity field was measured using scanning PIV \citep{Lawson2014}.
Tracer particles ($6~\mu\textrm{m}$ diameter polymethyl-methacrylate microspheres 
with specific gravity $1.22$) were seeded into the flow at a density of 
approximately 1 particle per $(1.4\eta)^3$ and illuminated by a $4.7\eta$ thick 
laser sheet, which was rapidly scanned across the measurement volume at $250~\textrm{Hz}$ 
using a galvanometer mirror scanner.
The tracers were observed in forward scatter orientation at $\pm 45^\circ$ to the 
laser sheet, imaged at a spatial resolution of $20 \ \mu \textrm{m}$ per pixel 
(1:2 optical magnification) by a pair of Phantom v640 high speed cameras 
at $15~\textrm{kHz}$. 
Each scan generated $54$ of $512\times 512$ pixel stereo image pairs,  
which were used to generate tomographic reconstructions of the tracer distribution.
Reconstructions from sequential scans were cross-correlated using a multi-pass 
PIV algorithm, with a final interrogation window size of $3.2\eta$, for an 
effective spatial resolution of around $1.6\eta$ and vector spacing $0.8\eta$.
This yielded 3D, 3-component velocity field measurements on a regular 
grid over a $(42\eta)^3$ measurement volume.

\paragraph*{\bf Evaluation of nonlinear term from CAV}

Taking the dot product of Eq.~\eqref{eq:vort} with $\ww$,
the equation for enstrophy $\Omega = |\ww|^2$ 
is given as
\begin{align}
\frac{1}{2} \frac{D \Omega}{D t} = (\ww \cdot \nabla \uu) \cdot \ww 
+ \nu \ww \cdot ( \nabla^2 \ww ) \ .
\label{eq:Om}
\end{align}
The amplification of enstrophy is engendered by
the nonlinear term $(\ww \cdot \nabla \uu) \cdot \ww $.
The conditional expectation 
$\langle (\ww \cdot \nabla \uu) \cdot \ww | \Omega \rangle$
allows us to assess the strength of nonlinear amplification
for a given enstrophy magnitude. 
It is more convenient to consider 
$ \langle (\hat{\ww} \cdot \nabla \uu) \cdot \hat{\ww} | \Omega \rangle
= \langle (\ww \cdot \nabla \uu) \cdot \ww | \Omega \rangle/ \Omega $,
since it provide the effective stretching, 
irrespective of the strength of the vorticity
\cite{Ashurst87, BBP2020}. 
As mentioned earlier, the amplification term 
can be solely obtained in terms of vorticity itself,
since the velocity field can be obtained from the vorticity
field using Eq.~\eqref{eq:biot}. 
Combined with conditional averaging, the effective stretching can be related to the CAV 
field introduced in Eq.~\eqref{eq:cav}.
To this end, we rewrite Eq.~\eqref{eq:biot}
using the index notation
\begin{align}
u_i (\xx) = \frac{1}{4\pi} \int \varepsilon_{ikl} \ \omega_l (\xx + \rr)
\ \frac{r_k}{r^3} \ d\rr \ , 
\label{eq:ui}
\end{align}
where $r = |\rr|$.
To evaluate the nonlinear term, we have to first obtain
the gradient of $\uu$. By taking the derivative
of Eq.~\eqref{eq:ui} w.r.t.~$x_j$ and using the theory
of singular integrals \cite{calderon65}, it can be 
shown that
\begin{align}
\begin{aligned}
\frac{\partial u_i}{\partial x_j} (\xx) 
&= - \frac{1}{2} \varepsilon_{ijk} \omega_k (\xx) + \\
& \frac{1}{4\pi}  \int 
\varepsilon_{ikl} \ \omega_l (\xx + \rr)  
\ \left[-\frac{\delta_{jk}}{r^3}  + 3 \frac{r_j r_k}{r^5} \right] \ d\rr  \ ,
\label{eq:uixj}
\end{aligned}
\end{align} 
where the integral on the right-hand side is taken in the sense
of principal value.

The nonlinear term now can obtained as 
\begin{align}
\begin{aligned}
\hat{\omega}_i \hat{\omega}_j \frac{\partial u_i}{\partial x_j} (\xx) 
= \frac{1}{4\pi} \int 
&\varepsilon_{ikl} \ \omega_l (\xx + \rr)
\hat{\omega}_i (\xx) \hat{\omega}_j (\xx) \\
& \ \left[-\frac{\delta_{jk}}{r^3}  + 3 \frac{r_j r_k}{r^5} \right] \ d\rr  \ .
\label{eq:wsw1}
\end{aligned}
\end{align} 
Using $\varepsilon_{ikl} \omega_i \omega_j \delta_{jk} = 0$
and 
with minor rearrangement, the expression then becomes
\begin{align}
\begin{aligned}
\hat{\omega}_i \hat{\omega}_j \frac{\partial u_i}{\partial x_j} (\xx) 
= \frac{3}{4\pi} \int 
&\hat{\omega}_i (\xx) \ \varepsilon_{ikl} 
\frac{r_k}{r} \omega_l (\xx + \rr) \\ 
& \ \frac{r_j}{r} \hat{\omega}_j (\xx) \ \frac{1}{r^3} \ d\rr \ .
\label{eq:wsw2}
\end{aligned}
\end{align} 
Reverting back to vector form, we can write
\begin{align}
\begin{aligned}
(\hat{\ww} \cdot \nabla \uu) \cdot \hat{\ww}
= \frac{3}{4\pi} \int 
&\hat{\ww} (\xx) \cdot 
\left[ \hat{\rr} \times \ww(\xx + \rr) \right] \\
&\  \frac{\hat{\rr} \cdot \hat{\ww} (\xx)}{r^3} \ d\rr  \ . 
\label{eq:wsw3}
\end{aligned}
\end{align} 
Without loss of generality, we can apply the conditioning
for the state $\zz = \ww(\xx)$
\begin{align}
\begin{aligned}
\langle (\hat{\ww} \cdot \nabla \uu) \cdot \hat{\ww} | \zz \rangle 
= \frac{3}{4\pi} \int 
& \hat{\zz} \cdot 
\left[ \hat{\rr} \times \langle \ww(\xx + \rr) | \zz \rangle \right] \\
& \  \frac{\hat{\rr} \cdot \hat{\zz}}{r^3} \ d\rr  \ . 
\label{eq:wsw4}
\end{aligned}
\end{align} 
Using $\langle \ww(\xx+\rr)|\zz\rangle$ from Eq.~\eqref{eq:cav},
together with 
$\hat{\zz} = \hat{\ww}(\xx) = \ee_z$ and 
$\hat{\rr} = \gamma \ee_z + (1-\gamma^2)^{1/2} \ee_\rho$,
it can be readily shown that 
\begin{align}
\begin{aligned}
\hat{\zz} \cdot (\hat{\rr} \times \langle \ww(\xx+\rr)|\zz\rangle) 
 = \overline{\omega}_\theta (1-\gamma^2)^{1/2} \ ; 
\end{aligned}
\end{align} 
whereas the term
$\hat{\rr} \cdot \zz = \hat{\rr} \cdot \ee_z = \gamma $, 
leading to the result
\begin{align}
\langle \hat{\ww} \cdot \nabla \uu \cdot \hat{\ww} | \zz \rangle =
\frac{3}{4\pi} \int 
\frac{\gamma ( 1- \gamma^2)^{1/2} \overline{\omega}_\theta}{r^3} d\rr \ . 
\label{eq:wsw_g}
\end{align}
Finally, a simple transformation to cylindrical
coordinates, with $\rho = r(1-\gamma^2)^{1/2}$, 
$z=r\gamma$ and $d\rr = \rho d\rho d\theta dz$,
with integration limits $\rho \in [0, \infty]$,
$\theta \in [0, 2\pi]$, $z \in [-\infty , \infty]$
gives the result shown in Eq.~\eqref{eq:wsw}.
Since $\omega_\theta$ does not depend on $\theta$
and is an odd function in $z$, 
we have integrated out $d\theta$ and appropriately 
changed the integration limits to $z \in [0, \infty]$.
Note that the scalar Eq.~\eqref{eq:wsw_g} is a function of the 
vorticity magnitude only. Therefore, we can equivalently choose the 
enstrophy as the conditioning variable in Eq.~\eqref{eq:wsw}.

\paragraph*{\bf Numerical evaluation of the 
CAV field:}

The CAV field defined in Eq.~\eqref{eq:cav}
can be extracted from experimental and DNS data
using two distinct methods. We note that we have
utilized both these methods, and the results
are essentially identical. Nevertheless, both
the methods have certain advantages and disadvantages,
so it is worth discussing them both.

\paragraph*{Binning and sampling approach.}
The first method entails straightforward conditional
averaging by simply accruing 
samples of vorticity vectors
at any two spatial locations. 
For instance, consider two spatial locations
$\xx_1=\xx$ and $\xx_2=\xx+\rr$ at time $t$, and the corresponding vorticity
vectors $\ww_1 = \ww(\xx_1,t)$ and $\ww_2 = \ww(\xx_2, t)$;
our task is to essentially evaluate
$\langle \ww_2 | \zz = \ww_1 \rangle$
for all combinations of the conditioning variable $\zz = \ww_1$ 
and the separation vector $\rr$.
As noted earlier, due to statistical isotropy,
the CAV field only depends on three scalar variables,
viz. $\Omega=|\zz|^2$, $r=|\rr|$ and 
$\gamma = \hat{\rr} \cdot \hat{\zz}$,
which can be readily calculated using $\ww_1$
and $\rr$. We can also obtain 
the orthogonal basis vectors 
$(\ee_z, \ee_\rho, \ee_\theta)$ as defined
immediately after Eq.~\eqref{eq:cav}. 
Thereafter, the components $\overline{\omega}_z, 
\overline{\omega}_\rho, \overline{\omega}_\theta$
are easily computed by taking simple dot products
of these unit vectors with the $\ww_2$ vector, i.e.,
$\overline{\omega}_{z,\rho, \theta} = \ee_{z, \rho, \theta} \cdot \ww_2$.
Thus, for each sample pair, we can obtain
the three scalar variables $\Omega$, $r$ and $\gamma$,
and also the three components $\overline{\omega}_{z, \rho , \theta}$.
The scalar variables
can be sampled into discrete bins and 
the components can be appropriately averaged
over all the samples to obtain the conditional
expectations 
$\langle \overline{\omega}_{z, \rho, \theta} |\Omega, r, \gamma\rangle $. 
The CAV field itself can then be readily reconstructed
from these components for any choice of conditioning vorticity
and separation vector as necessary.

Evidently, the above method is
simple and straightforward in application.
By taking two spatial points at a time, 
and scanning through all such samples, the conditional
averages can be obtained very easily. 
However, for a $N^3$ grid, the total
number of pairs that can be formed scales as 
$\mathcal{O}(N^6)$,
which is obviously prohibitively expensive. 
Thus, only a limited number of pairs can be counted
using this method. Additionally complexities
arise in data processing, since the domain
is distributed across multiple processors,
see e.g. \cite{BSY.2015, BYS.2016} for 
an exposition on similar issues arising in the context
of processing pairs of particle trajectories.
Thus, for convenience, we only process pairs of samples
along given Cartesian grid lines. Even then the computational
cost can be very high, since for $N$ points along a grid line,
there are $\mathcal{O}(N^2)$ samples, implying a total
of $\mathcal{O}(N^4)$ samples for the entire grid. 
Since, we are primarily concerned with the localized structure
around intense vorticity, we restrict the samples
to typically $r\lesssim 100\eta$
(which still amounts to about $100-1000 N^3$ samples).

\paragraph*{Spectral approach.}
The difficulty of the first approach can be mitigated
by using a spectral approach to directly evaluate the entire CAV field.
To that end,  we can rewrite the conditional expectation in 
Eq.~\eqref{eq:cav0} as 
\begin{align}
\langle \ww (\xx+\rr,t) |\zz \rangle 
=  \frac{1}{f(\zz)} \langle \ww (\xx + \rr,t)  f^\prime (\zz ; \xx, t) \rangle \ ,
\label{eq:cav_fg}
\end{align}
where $f^\prime (\zz ; \xx, t) = \delta (\ww(\xx,t) - \zz)$
is the fine-grained probability density function (PDF)
of vorticity, and $f (\zz) = \langle f^\prime (\zz) \rangle $ 
is simply the PDF of vorticity. For details
pertaining to above relation see e.g., Appendix H in \cite{popebook}.  
Thus, the CAV field can be obtained as a convolution
between then vorticity field and the fine-grained
PDF.  This convolution can be evaluated
spectrally for any chosen conditioning vorticity as follows.
First, we define the indicator function
$I(\xx) = \delta (\ww (\xx,t) - \zz)$, which is set to unity
at every grid point 
where the vorticity condition is satisfied, and zero everywhere else.
Note that to satisfy the condition, 
both the vorticity magnitude and orientation must be matched. 
For the magnitude, we simply consider the same finite bins for
enstrophy as utilized before (in the binning approach), 
whereas for the orientation, we consider vorticity aligned
with any Cartesian grid direction, within 
a small tolerance (of about 5\%). 
Thereafter, a Fourier transform of the vorticity field
and the indicator function is performed. 
The Fourier transform
of the CAV field can be defined from 
Eq.~\eqref{eq:cav_fg} as 
\begin{align}
\tilde{\overline{\ww}}(\kk, \zz) = 
\frac{1}{f(\zz)} 
\langle \tilde{\ww}^*(\kk) \tilde{I} (\kk) \rangle \ , 
\end{align}
where $(\hat{\cdot})$ denotes the Fourier
coefficient and $\kk$ is the wave-vector. 
Thus, the CAV field can be obtained in spectral space 
by performing a convolution of the vorticity and indicator function,
and finally in physical space by taking an inverse transform.


Evidently, the main advantage of the spectral approach is that
it provides a direct evaluation of the CAV field in the entire
domain at the same cost as a 3D-FFT, which is $N^3\log_2 N$,
which is obviously more cost-effective than the earlier binning
approach. 
However, it also only provides the CAV field for one 
chosen conditional
magnitude at a time, i.e., the entire procedure has to be repeated 
for every conditional magnitude desired. 
Since the number of chosen conditional values is fixed
and does not depend on $N$,  the overall
cost for this approach still scales as $N^3 \log_2 N$. 
As mentioned earlier, we have utilized both the approaches,
and they essentially give identical results (as expected).
However, the spectral approach is particularly useful in setting
up the initial condition in Fig.~\ref{fig:cav_lag}, since it
needs to be performed only for one chosen value of conditional enstrophy,
and also for the entire domain.

\paragraph*{\bf Initial condition for dynamical evolution.}

The vortex structure which serves as the initial condition
in Fig.~\ref{fig:cav_lag}
can be extracted in two ways. For the first approach,
the CAV field is extracted at an earlier time,
at which point the conditional vorticity is
weak and still undergoing amplification. 
This can be done
by generalizing the CAV framework to two times, i.e.
$\langle \ww (\xx+\rr, t_1) | \ww (\xx, t_0)\rangle$, 
where $t_1 < t_0$, and $t_0$ corresponds
to the time at which maximum enstrophy is reached. 
In this approach, one requires Lagrangian trajectories
stored with the Eulerian data in DNS,
which can be expensive.

The second approach is to utilize the 
CAV field as obtained earlier, and
briefly evolve it backwards in time until
an initial condition is obtained where 
the vorticity is relatively weak and 
only the regular twisting of vortex lines
exist. This approach is far more efficient
for implementation, and hence we utilize this here.
However, we note that we have also verified 
using the first approach and both of them
essentially give the same outcome (with
some minor quantitative differences).


\section*{Acknowledgements}

D.B. gratefully acknowledges the Gauss Centre for Supercomputing e.V.
(www.gauss-centre.eu) for providing computing time on the
GCS supercomputers JUQUEEN and JUWELS at J\"ulich Supercomputing Centre (JSC), 
where the simulations reported in this paper were performed. 
M.W. acknowledges support by the European Research Council (ERC) 
under the European Union's Horizon 2020 research and 
innovation programme (Grant agreement No. 101001081).
J.M.L. acknowledges the IRIDIS High Performance Computing 
Facility and associated support services at the University of Southampton.
We thank T. D. Drivas for comments and suggesting
some pertinent references.

\section*{Author contributions}

D.B. performed numerical simulations. J.M.L. performed experiments.
All authors designed the research and discussed the results.
D.B. wrote the manuscript, D.B. and M.W. reviewed and made edits,
J.M.L. provided comments.

\section*{Competing interests} 

The authors declare no competing  interests.

\section*{Data and materials availability}

All data needed to support the conclusions are included in the article.

\bibliographystyle{unsrt}

\begin{thebibliography}{10}

\bibitem{she_90}
Z.-S. She, E.~Jackson, and S.~A. Orszag.
\newblock Intermittent vortex structures in homogeneous isotropic turbulence.
\newblock {\em Nature}, 344(6263):226, 1990.

\bibitem{polanco21}
J.~I. Polanco, N.~P. M{\"u}ller, and G.~Krstulovic.
\newblock Vortex clustering, polarisation and circulation intermittency in
  classical and quantum turbulence.
\newblock {\em Nat. Commun.}, 12:7090, 2021.

\bibitem{kleckner2013}
D.~Kleckner and W.~T~M. Irvine.
\newblock Creation and dynamics of knotted vortices.
\newblock {\em Nature Phys.}, 9:253--258, 2013.

\bibitem{zwierlein}
M.~W. Zwierlein, J.~R. Abo-Shaeer, A.~Schirotzek, C.~H. Schunck, and
  W.~Ketterle.
\newblock Vortices and superfluidity in a strongly interacting fermi gas.
\newblock {\em Nature}, 435:1047--1051, 2005.

\bibitem{freilich2010}
D.~V. Freilich, D.~M. Bianchi, A.~M. Kaufman, T.~K. Langin, and D.~S. Hall.
\newblock Real-time dynamics of single vortex lines and vortex dipoles in a
  {Bose-Einstein} condensate.
\newblock {\em Science}, 329:1182--1185, 2010.

\bibitem{aharon}
A.~Aharon-Steinberg, T.~Voelkl, A.~Kaplan, A.~Pariari, I.~Roy, T.~Holder,
  Y.~Wolf, A.~Suhov, Y.~Myasoedov, M.~Huber, B.~Yan, G.~Falkovich, L.~Levitov,
  M.~Hücker, and E.~Zeldov.
\newblock Direct observation of vortices in an electron fluid.
\newblock {\em Nature}, 607:74--80, 2022.

\bibitem{falkovich02}
G.~Falkovich, A.~Fouxon, and M.~G. Stepanov.
\newblock Acceleration of rain initiation by cloud turbulence.
\newblock {\em Nature}, 419:151, 2002.

\bibitem{dritschel}
D.~G. Dritschel and B.~Legras.
\newblock Modeling oceanic and atmospheric vortices.
\newblock {\em Physics Today}, 46:44--51, 1993.

\bibitem{ahmed2012}
H.~E. Ahmed, H.~A. Mohammed, and M.~Z. Yusoff.
\newblock An overview on heat transfer augmentation using vortex generators and
  nanofluids: approaches and applications.
\newblock {\em Renew. Sustain. Energy Rev.}, 16:5951--5993, 2012.

\bibitem{spalart1998}
P.~R. Spalart.
\newblock Airplane trailing vortices.
\newblock {\em Annu. Rev. Fluid Mech.}, 30:107, 1998.

\bibitem{pletcher}
R.~H. Pletcher, J.~C. Tannehill, and D.~Anderson.
\newblock {\em Computational fluid mechanics and heat transfer}.
\newblock CRC press, 1997.

\bibitem{Fefferman}
C.~Fefferman.
\newblock {\em Existence and smoothness of the {Navier-Stokes} equations}.
\newblock Clay Mathematical Institute, Cambridge, MA, 2006.

\bibitem{doering2009}
C.~R. Doering.
\newblock {The {3D} {Navier-Stokes} problem}.
\newblock {\em Annu. Rev. Fluid Mech.}, 41:109--128, 2009.

\bibitem{tl72}
H.~Tennekes and J.~L. Lumley.
\newblock {\em A First Course in Turbulence}.
\newblock The MIT Press, Cambridge, Massachussets and London, England, 1972.

\bibitem{Tsi2009}
A.~Tsinober.
\newblock {\em An Informal Conceptual Introduction to Turbulence}.
\newblock Springer, Berlin, 2009.

\bibitem{BBP2020}
D.~Buaria, E.~Bodenschatz, and A.~Pumir.
\newblock Vortex stretching and enstrophy production in high {Reynolds} number
  turbulence.
\newblock {\em Phys. Rev. Fluids}, 5:104602, 2020.

\bibitem{beale84}
J.~T. Beale, T.~Kato, and A.~Majda.
\newblock Remarks on the breakdown of smooth solutions for the {3-D Euler}
  equations.
\newblock {\em Commun. Math. Phys.}, 94:61--66, 1984.

\bibitem{Jimenez93}
J.~Jim{\'e}nez, A.~A. Wray, P.~G. Saffman, and R.~S. Rogallo.
\newblock The structure of intense vorticity in isotropic turbulence.
\newblock {\em J. Fluid Mech.}, 255:65--90, 1993.

\bibitem{Ishihara09}
T.~Ishihara, T.~Gotoh, and Y.~Kaneda.
\newblock Study of high-{Reynolds} number isotropic turbulence by direct
  numerical simulations.
\newblock {\em Annu. Rev. Fluid Mech.}, 41:165--80, 2009.

\bibitem{BPBY2019}
D.~Buaria, A.~Pumir, E.~Bodenschatz, and P.~K. Yeung.
\newblock Extreme velocity gradients in turbulent flows.
\newblock {\em New J. Phys.}, 21:043004, 2019.

\bibitem{Saw:2016}
E.-W. Saw, D.~Kuzzay, D.~Farande, A.~Guittonneau, F.~Daviaud,
  C.~Wiertel-Gasquet, V.~Padilla, and B.~Dubrulle.
\newblock Experimental characterization of extreme events of inertial
  dissipation in a turbulent swirling flow.
\newblock {\em Nature Comm.}, 7:12566, 2016.

\bibitem{tao2019}
T.~Tao.
\newblock {Searching for singularities in the {Navier-Stokes} equations}.
\newblock {\em Nature Rev. Phys.}, 1:418--419, 2019.

\bibitem{BPB2020}
D.~Buaria, A.~Pumir, and E.~Bodenschatz.
\newblock Self-attenuation of extreme events in {Navier-Stokes} turbulence.
\newblock {\em Nat. Commun.}, 11:5852, 2020.

\bibitem{ham_pof08}
P.~E. Hamlington, J.~Schumacher, and W.~J.~A. Dahm.
\newblock Direct assessment of vorticity alignment with local and nonlocal
  strain rates in turbulent flows.
\newblock {\em Phys. Fluids}, 20:111703, 2008.

\bibitem{BP2021}
D.~Buaria and A.~Pumir.
\newblock Nonlocal amplification of intense vorticity in turbulent flows.
\newblock {\em Phys. Rev. Research}, 3:042020, 2021.

\bibitem{constantin1993}
P.~Constantin and C.~Fefferman.
\newblock Direction of vorticity and the problem of global regularity for the
  {N}avier-{S}tokes equations.
\newblock {\em Indiana University Mathematics Journal}, 42:775--789, 1993.

\bibitem{constantin1996}
P.~Constantin, C.~Fefferman, and A.~J. Majda.
\newblock Geometric constraints on potentially singular solutions for the {3-D
  Euler} equations.
\newblock {\em Commun. Partial. Differ. Equ.}, 21(3-4), 1996.

\bibitem{kerr:1993}
R.~M. Kerr.
\newblock Evidence for a singularity of the three-dimensional, incompressible
  {Euler} equations.
\newblock {\em Phys. Fluids A}, 5:1725--1746, 1993.

\bibitem{luo14}
G.~Luo and T.~Y. Hou.
\newblock Potentially singular solutions of the {3D} axisymmetric {E}uler
  equations.
\newblock {\em Proc. Nat. Acad. Sci.}, 111:12968--12973, 2014.

\bibitem{novikov93jfr}
E.~A. Novikov.
\newblock A new approach to the problem of turbulence, based on the
  conditionally averaged {N}avier-{S}tokes equations.
\newblock {\em Fluid Dyn. Res.}, 12(2):107 -- 126, 1993.

\bibitem{mui1996}
R.~C.~Y. Mui, D.~G. Dommermuth, and E.~A. Novikov.
\newblock Conditionally averaged vorticity field and turbulence modeling.
\newblock {\em Phys. Rev. E}, 53:2355, 1996.

\bibitem{wilczek_thesis}
M.~Wilczek.
\newblock {\em Statistical and numerical investigations of fluid turbulence}.
\newblock PhD thesis, Westf{\"a}lische Wilhelms-Universit{\"a}t M{\"u}nster,
  2011.

\bibitem{Friedrich:2012}
R.~Friedrich, A.~Daitche, O.~Kamps, J.~L{\"u}lff, M.~Vo{\ss}kuhle, and
  M.~Wilczek.
\newblock The {L}undgren-{M}onin-{N}ovikov hierarchy: kinetic equations for
  turbulence.
\newblock {\em C. R. Phys.}, 13:929--953, 2012.

\bibitem{Lawson2019}
J.~M. Lawson, E.~Bodenschatz, A.~N. Knutsen, J.~R. Dawson, and N.~A. Worth.
\newblock {Direct assessment of Kolmogorov's first refined similarity
  hypothesis}.
\newblock {\em Phys. Rev. Fluids}, 4:022601(R), 2019.

\bibitem{Lawson2014}
J.~M. Lawson and J.~R. Dawson.
\newblock {A scanning PIV method for fine-scale turbulence measurements}.
\newblock {\em Expt. Fluids}, 55:1857, 2014.

\bibitem{mm98}
P.~Moin and K.~Mahesh.
\newblock Direct numerical simulation: a tool in turbulence research.
\newblock {\em Annu. Rev. Fluid Mech.}, 30:539--578, 1998.

\bibitem{Frisch95}
U.~Frisch.
\newblock {\em Turbulence: the legacy of {Kolmogorov}}.
\newblock Cambridge University Press, Cambridge, 1995.

\bibitem{BP2022}
D.~Buaria and A.~Pumir.
\newblock Vorticity-strain rate dynamics and the smallest scales of turbulence.
\newblock {\em Phys. Rev. Lett.}, 128:094501, 2022.

\bibitem{BS_PRL_2022}
D.~Buaria and K.~R. Sreenivasan.
\newblock Scaling of acceleration statistics in high {Reynolds} number
  turbulence.
\newblock {\em Phys. Rev. Lett.}, 128:234502, 2022.

\bibitem{BS_PRL_2023}
D.~Buaria and K.~R. Sreenivasan.
\newblock Saturation and multifractality of {Lagrangian and Eulerian} scaling
  exponents in three-dimensional turbulence.
\newblock {\em Phys. Rev. Lett.}, 131:204001, 2023.

\bibitem{Burgers48}
J.~M. Burgers.
\newblock A mathematical model illustrating the theory of turbulence.
\newblock {\em Adv. Appl. Mech.}, 1:171--99, 1948.

\bibitem{Betchov56}
R.~Betchov.
\newblock An inequality concerning the production of vorticity in isotropic
  turbulence.
\newblock {\em J. Fluid Mech.}, 1:497--504, 1956.

\bibitem{gibbon08}
J.~D. Gibbon.
\newblock The three-dimensional {E}uler equations: Where do we stand?
\newblock {\em Physica D}, 237:1894--1904, 2008.

\bibitem{elgindi2021}
T.~M. Elgindi.
\newblock Finite-time singularity formation for $\mathbb{C}^{1,\alpha}$
  solutions to the incompressible {E}uler equations on $\mathbb{R}^3$.
\newblock {\em Ann. Math.}, 194:647--727, 2021.

\bibitem{Ashurst87}
W.~T. Ashurst, A.~R. Kerstein, R.~M. Kerr, and C.~H. Gibson.
\newblock Alignment of vorticity and scalar gradient with strain rate in
  simulated {Navier-Stokes} turbulence.
\newblock {\em Phys. Fluids}, 30:2343--2353, 1987.

\bibitem{BPB2022}
D.~Buaria, A.~Pumir, and E.~Bodenschatz.
\newblock Generation of intense dissipation in high {Reynolds} number
  turbulence.
\newblock {\em Philos. Trans. R. Soc. A}, 380:20210088, 2022.

\bibitem{BP2023}
D.~Buaria and A.~Pumir.
\newblock Role of pressure in the dynamics of intense velocity gradients in
  turbulent flows.
\newblock {\em J. Fluid Mech.}, 973:A23, 2023.

\bibitem{grafke2015}
T.~Grafke, R.~Grauer, and T.~Schäfer.
\newblock The instanton method and its numerical implementation in fluid
  mechanics.
\newblock {\em J. Phys. A Math. Theor.}, 48:333001, 2015.

\bibitem{schorlepp22prsa}
T.~Schorlepp, T.~Grafke, S.~May, and R.~Grauer.
\newblock Spontaneous symmetry breaking for extreme vorticity and strain in the
  three-dimensional {N}avier–{S}tokes equations.
\newblock {\em Phil. Trans. R. Soc. A}, 380(2226):20210051, 2022.

\bibitem{melander94}
M.~V. Melander and F.~Hussain.
\newblock Core dynamics on a vortex column.
\newblock {\em Fluid Dyn. Res.}, 13:1, 1994.

\bibitem{EP88}
V.~Eswaran and S.~B. Pope.
\newblock An examination of forcing in direct numerical simulations of
  turbulence.
\newblock {\em Comput. Fluids}, 16:257--278, 1988.

\bibitem{Rogallo}
R.~S. Rogallo.
\newblock Numerical experiments in homogeneous turbulence.
\newblock {\em NASA Technical Memo}, 1981.

\bibitem{PattOrs71}
G.~S. Patterson and S.~A. Orszag.
\newblock Spectral calculations of isotropic turbulence: efficient removal of
  aliasing interactions.
\newblock {\em Phys. Fluids}, 14:2538--2541, 1971.

\bibitem{BS2022}
D.~Buaria and K.~R. Sreenivasan.
\newblock Intermittency of turbulent velocity and scalar fields using
  three-dimensional local averaging.
\newblock {\em Phys. Rev. Fluids}, 7:L072601, 2022.

\bibitem{BS2023}
D.~Buaria and K.~R. Sreenivasan.
\newblock {Lagrangian acceleration and its Eulerian decompositions in fully
  developed turbulence}.
\newblock {\em Phys. Rev. Fluids}, 8:L032601, 2023.

\bibitem{BSY.2015}
D.~Buaria, B.~L. Sawford, and P.~K. Yeung.
\newblock Characteristics of backward and forward two-particle relative
  dispersion in turbulence at different {R}eynolds numbers.
\newblock {\em Phys. Fluids}, 27:105101, 2015.

\bibitem{Xu2007}
H.~Xu, N.~T. Ouellette, D.~Vincenzi, and E.~Bodenschatz.
\newblock {Acceleration Correlations and Pressure Structure Functions in
  High-Reynolds Number Turbulence}.
\newblock {\em Phys. Rev. Lett.}, 99:204501, 2007.

\bibitem{calderon65}
A.~P. Calder{\'o}n and A.~Zygmund.
\newblock {\em Singular integrals}.
\newblock American Mathematical Society, 1965.

\bibitem{BYS.2016}
D.~Buaria, P.~K. Yeung, and B.~L. Sawford.
\newblock {A Lagrangian} study of turbulent mixing: forward and backward
  dispersion of molecular trajectories in isotropic turbulence.
\newblock {\em J. Fluid Mech.}, {799}:{352--382}, 2016.

\bibitem{popebook}
S.~B. Pope.
\newblock {\em Turbulent Flows}.
\newblock Cambridge University Press, 2000.

\end{thebibliography}

\end{document}